\documentstyle[aps,pra,epsf]{revtex} 
\begin{document}
\draft
\tighten
\onecolumn
\title{\bf Decoherence and dephasing in spin-based  
solid state quantum computers}
\author{Xuedong Hu, Rogerio de Sousa, and S. Das Sarma}
\address{Department of Physics, University of Maryland, College Park, 
MD 20742-4111}
\date{\today}
\maketitle
\begin{abstract}
We discuss the operational definition of decoherence in various
solid state systems.  In particular, we review, in the
context of spin-based solid state quantum computation, 
the introduction of $T_1$ and $T_2$ to describe decoherence in a two
level system.  We provide a perspective on
recent experiments involving the manipulation of spin coherence in
semiconductors, and discuss specific decoherence and dephasing issues 
in electron spin-based quantum dot quantum computer architectures.
\end{abstract}

\section{Introduction}

A quantum eigenstate of a particular Hamiltonian is by definition a
stationary state, in which the wavefunction might vary spatially, but
does not decay in time.  A quantum system can be in a
superposition of its eigenstates with definite phase and
amplitude relationships among the basis states.  Such a
superposition of eigenstates with definite phase relationships is
called quantum coherence.  Decoherence refers loosely to how a system
loses these quantum coherence features.  For example, it can refer to
the amplitude decay (often exponential) and the associated
disappearance of a quantum eigenstate in time (by virtue of it
interacting with a surrounding bath, for instance).  Or it may refer
to the loss of electron phase coherence, because the definite phase
relationship between the superposing states disappears over time
leading to dephasing.  

In an isolated quantum system decoherence could only arise from the
dynamical degrees of freedom neglected in the original Hamiltonian
used to define the quantum state.  In a system coupled to an external
bath, decoherence could arise naturally from the coupling between the
system and the bath as the quantum eigenstate (presumably slowly)
leaks to the environment (the bath) due to energy exchange between
the system and the bath.  Note that the bath or the environment does
not need to be physically separated from the ``system''---it is a
standard practice in physics to divide a large system into
sub-systems which are ``reasonably isolated'' (in some well-defined
operational sense) from each other, i.e. the interaction
Hamiltonian coupling the various sub-systems is ``weak'' in a
precisely defined manner.  In situations like this (which are
commonplace in condensed matter physics) the ``system'' and the
``bath'' can be two components of the same system (such as electrons
and phonons of a crystal lattice).

Decoherence in quantum mechanics
has received a great deal of recent attention in the context of
current interest in quantum computation and information processing. 
In a quantum computer (QC), an algorithm is typically performed by
applying unitary operations on an array of two level systems (qubits)
that carry quantum information \cite{nielsen}.  For quantum
computation, these qubits must be isolated from the other degrees of
freedom that might perturb their unitary time evolution.  In other
words, decoherence in a quantum computer must be much slower than a
typical quantum gate operation for successful quantum computation. 
The ratio between gate time and decoherence time needs to be smaller
than $10^{-3} \sim 10^{-6}$ \cite{nielsen}, which is the current
quantum error correction limit.  Thus the control of decoherence
is a crucial aspect of quantum information processing. 
Decoherence is not only pertinent to performing
quantum computation but is also a central issue in describing the
transition between regimes governed by quantum mechanics and
classical mechanics \cite{zurek}.  In fact, decoherence is a
fundamental concept in the quantum measurement theory and in works
dealing with the foundations of quantum mechanics.

In condensed matter physics, terminologies such as decoherence,
dephasing, relaxation, scattering, etc, are often used quite liberally
(and somewhat confusingly)
due to the diversity of physical systems and
phenomena studied.  With the many existing proposals for solid state
quantum computer architectures [4-33], there is obviously a need to
accurately understand decoherence issues in each particular scheme. 
The specific decoherence channels are always system dependent, but
there also exist many common features.  For example, most QC proposals
involve quantum two level systems (TLS) \cite{cohen} serving the role
of  qubits.  These TLS might be electron spin levels, electron
orbital levels, nuclear spin levels, charging levels, magnetic flux
directions, etc.  The basic QC algorithm in each scheme
involves dynamic manipulations of these TLS using various
external means in order to perform one and two qubit
operations.  It is therefore imperative that the decoherence
time in these TLS dynamics is much larger than the qubit operation
times.  Because of the two-level nature of these systems, it is
possible to describe their decoherence using just two dephasing time
scales: namely $T_{1}$ and $T_{2}(\le T_{1})$, which give a
phenomenological description of the population and phase relaxation
in these systems.  For an ensemble of TLS, another time scale
$T_{2}^{*}\le T_{2}$ should also be defined, since some
spins may rotate faster than others leading to reversible loss of
quantum coherence between them.  These two time scales ($T_2^*$ and
$T_2$) should be carefully distinguished, as the current QC proposals
mostly involve single TLS so that $T_2$ is the relevant quantity,
while in macroscopic measurements the observed quantity is often
$T_2^*$ due to the ensemble averaging over the response from a large
number of TLS.  We should note that using just two relaxation times
almost never gives a complete description of the dynamics and
lineshapes of a realistic two-level system coupled to a physical
environment.  However, in several paradigmatic TLS such as nuclear
spins probed by NMR (nuclear magnetic resonance) or electron spins
probed by ESR (electron spin resonance), relaxation constants 
$T_{1}$ and $T_{2}$ give the proper qualitative description
of the signal linewidths, and are good operational representations of
the various relaxation channels.  We also note that in many
situations of interest $T_{1}$ and $T_{2}$ could be equal (or quite
close in values).

In this paper we first review and discuss the definitions and
significance of $T_1$, $T_2$, and $T_2^*$ in the context of 
macroscopic Bloch equations and the density matrix representation. 
We then review the role of these relaxation times in experiments
involving semiconductors, which are projected to be of potential use
in several proposed solid state QC architectures.  We also
discuss some specific features of decoherence in a solid state
quantum computer, particularly aspects of gate induced decoherence in
the form of non-adiabaticity, electrical circuit noise, and field
inhomogeneity. We conclude with a brief discussion of electronic
decoherence in mesoscopic systems.

\section{Decoherence and the Bloch equations}

As decoherence occurs in the time domain, it is natural to use
time scales to describe how strong a decoherence channel is, and
how strongly decoherence affects a particular dynamic variable.
If we restrict ourselves to a simple single TLS, the problem of 
describing phenomenologically the effect of weak coupling to
external degrees of freedom on the time evolution of this TLS
contains only a few parameters.  In particular, 
two time scales of relaxation, $T_1$ and $T_2$, were
introduced and used extensively in the field of 
NMR \cite{slichter}, and then used naturally also in
ESR \cite{Poole} and in quantum optics
\cite{meystre}, in which either the applied static magnetic field
(which causes Zeeman splitting along the field direction) or the
natural TLS (e.g. for photons, where the longitudinal and transverse
polarization occurs naturally) defines a longitudinal and a transverse
direction.  $T_1$ and $T_2$ are then respectively the longitudinal and
transverse relaxation times for magnetizations in NMR and ESR, or the
population difference and polarization in quantum optics.  Note that
using $T_1$ and $T_2$ to characterize decoherence
applies only to TLS dynamics.

The definition of $T_1$ and $T_2$ is quite system specific---in
fact, the strict definition of $T_1$ and $T_2$ apply specifically
to magnetic resonance measurements.  An arbitrary decoherence
phenomenon might require  more or less parameters to describe the
dephasing process.  In general, in the absence of magnetic field
and in isotropic systems, $T_1 = T_2$.  In weak
localization related mesoscopic experiments, a single electron
dephasing time, which parametrizes the loss of quantum interference,
is sufficient to characterize the results. We would also like to point
out that $T_1$ and $T_2$ are purely phenomenological parameters
(characterizing longitudinal and transverse relaxation respectively),
to which many different decoherence mechanisms could, in principle,
contribute.  In general, two parameters may not be adequate to
completely describe TLS decoherence, but experience (particularly in
NMR, ESR, and optical pumping experiments) suggests that $T_1$ and
$T_2$ are often quite sufficient in characterizing TLS decoherence in
many diverse situations and are therefore extremely important TLS
parameters.

We first discuss how $T_1$ and $T_2$ are introduced for a particular
TLS---an electron spin in an external magnetic field.
Consider an electron subject to a constant field in the z direction 
${\mathbf B_{\parallel}}=B_{\parallel} \; {\mathbf {\hat z}}$,
and a rotating field in the xy plane with frequency $\nu$ which 
perturbs its spin precession,
${\mathbf B_{\perp}}=B_{\perp} [\cos (\nu t){\mathbf {\hat x}} + 
\sin(\nu t) {\mathbf {\hat y}}]$. The magnetic moment of the
electronic spin is given by  
\begin{equation}
{\mathbf {\vec \mu}}=\frac{ge}{2mc}{\mathbf S} = -\frac{g}{2}\mu_{e}
\mathbf{{\vec \sigma}},
\end{equation}
where $\mu_{e}$ is the Bohr magneton, $g$ is the electron
gyromagnetic ratio, and ${\mathbf {\vec
\sigma}}=\sigma_{x}{\mathbf {\hat x}}+\sigma_{y}{\mathbf {\hat
y}}+\sigma_{z}{\mathbf {\hat z}}$ is the Pauli spin vector. The
spin Hamiltonian $H=-{\mathbf {\vec \mu}} \cdot {\mathbf B}$ is
given by its parallel and perpendicular parts,
$H=H_{\parallel}+H_{\perp}$: 
\begin{equation}
H = - {\vec \mu} \cdot (B_{\perp} [\cos (\nu t){\mathbf {\hat x}} + 
\sin(\nu t) {\mathbf {\hat y}}] + B_{\parallel} \, {\mathbf {\hat z}}) 
=\frac{1}{2}
\left(
\begin{array}{cc}
\hbar \omega & 2\Delta e^{-i \nu t}\\
2\Delta e^{i\nu t} & -\hbar \omega \\
\end{array}
\right),
\label{htot}
\end{equation}
where $\hbar \omega = g \mu_e B_{\parallel}$ and
$\Delta=\frac{g}{2}\mu_{e}B_{\perp}$.
%
%It is worth noting that if the oscillating field is linear, say in 
%the x direction ${\mathbf B_{\perp}^{L}}=B_{L}\cos(\nu t){\mathbf
%{\hat x}}$, then the rotating wave approximation \cite{meystre}
%asserts that a rotating field in the xy plane 
%is a very good approximation to the linear field solution.
%
The spin dynamics under
Hamiltonian (\ref{htot}) is governed by the von Neumann
equation for the density matrix, 
\begin{equation} 
i\hbar {\dot \rho}=[H,\rho].
\end{equation}
We can remove the time dependence of $H$ by going to the
interaction picture rotating in phase with ${\mathbf
B_{\perp}}$, in which
the equation of motion is much simplified. 
Since $\rho_{\uparrow \uparrow}+\rho_{\downarrow \downarrow}=1$ and 
$\rho_{\uparrow \downarrow}=\rho_{\downarrow \uparrow}^{*}$, only two
equations describe the dynamics in the interaction picture:
%
%(we note that
%in the interaction picture the
%off-diagonal term $e^{i \nu t}$ is factored out):  
%
\begin{eqnarray}
i\hbar {\dot \rho}_{\uparrow \uparrow}&=&\Delta (\rho_{\uparrow 
\downarrow}^{*}-\rho_{\uparrow \downarrow}), 
\label{evo1}\\
i\hbar {\dot \rho}_{\uparrow \downarrow}&=&\hbar \delta \rho_{\uparrow 
\downarrow}+\Delta (\rho_{\downarrow \downarrow}-\rho_{\uparrow
\uparrow}). 
\label{evo2}
\end{eqnarray}
where $\delta=\omega - \nu$ is the detuning frequency of the applied
transverse rotating field. 

The evolution of this rotating or precessing electron spin is unitary
since up to now we are considering a single isolated spin without any
dephasing or decoherence.  Quantum coherence is always
maintained.  However, an electron spin is never
isolated.  It couples to the electron orbital degrees of freedom
through spin-orbit coupling, to nuclear spins through hyperfine
interaction and dipole interaction, to the crystal lattice (and thus
phonons) through spin-orbit coupling, to any magnetic impurities in
the environment through direct spin dipole coupling, and to other
electron spins through dipole and exchange coupling.  With
all these ``environmental'' degrees of freedom being present in
principle, the simple decoherence-free equations for the spin density
matrix above are obviously an idealization.  To determine how strong
these ``outside'' influences are, we need to include them in the
starting Hamiltonian, and then employ approaches such as master
equations or Langevin equations in order to achieve a quantitative
understanding of how the system evolves and how the spin we are
concerned with loses its coherence due to its coupling to the
``bath'', which, in this case, implies the collection of all the
degrees of freedom neglected in our Hamiltonian of Eq.~(\ref{htot}). 
This is obviously a very complicated problem (in fact, an insoluble
problem since we can never be sure of all the possible degrees of
freedom in the ``bath'').  A simple approach to tackle this problem
is to add exponential decay terms to the right hand sides of the two
equations above: $-\frac{i\hbar}{T_{1}}\rho_{\uparrow \uparrow}$ to
Eq. (\ref{evo1}) and $-\frac{i\hbar}{T_{2}}\rho_{\uparrow
\downarrow}$ to Eq. (\ref{evo2}) to mimic decoherence
phenomenologically.  This is similar to adding a friction term
proportional to velocity in the classical Newton's equation, and is
essentially a Markovian approximation to spin relaxation as it
neglects memory effects in the system-environment coupling.  The two
time constants can be calculated under certain conditions if
sufficient information about the bath is available.  This simple
phenomenological approach turns out to be quite successful in
describing many experiments, ranging from NMR and ESR to quantum
optics, although actual explicit calculations of $T_1$ and $T_2$ are
generally quite difficult.   It is perhaps more fruitful to consider
$T_1$ and $T_2$ as purely phenomenological parameters (characterizing
longitudinal and transverse relaxations respectively in the TLS
dynamics) to be obtained from experimental measurements.

Recall that the unit magnetization vector is related to the spin
via ${\mathbf M}=Tr(\rho_{I}{\mathbf {\vec \sigma}})$,
so that $M_{x}=2Re(\rho_{\uparrow \downarrow})$,
$M_{y}=-2Im(\rho_{\uparrow \downarrow})$, $M_{z}=(\rho_{\uparrow
\uparrow}-\rho_{\downarrow \downarrow})$.  We can now rewrite Eqs.
(\ref{evo1}) and (\ref{evo2}) with respect to the components of the
real vector ${\mathbf M}$, obtaining the \textit{Bloch
equations} in the interaction picture \cite{meystre,bloch}:  
\begin{eqnarray} 
{\dot M}_{x} &=&
-\frac{1}{T_{2}}M_{x}-\delta M_{y}, \\ 
{\dot M}_{y} &=& \delta
M_{x}-\frac{1}{T_{2}}M_{y}+R_{0}M_{z}, \\ 
{\dot M}_{z} &=& -R_{0}M_{y} - \frac{1}{T_{1}}(M_{z}+1), 
\end{eqnarray} 
where
$R_{0}=-2\Delta/\hbar=-g e_{0}B_{\perp}/(mc)$ is the Rabi flopping
frequency, negative for a free electron, whose magnetization rotates
counter-clockwise in the yz plane (we keep this sign for conventional
purposes).  We readily see that $T_{2}$ is
the relaxation time for the xy (transverse) magnetization of the
electrons, whereas $T_{1}$ is the decay time for the z direction
(longitudinal) magnetization.  To describe an ensemble of spins, which
may in general possess different Zeeman splittings $\hbar \omega$ (for
example, by virtue of inhomogeneities in the applied magnetic field
and/or in the electron g-factor) and thus having different detunings
$\delta$ in the rotating field, additional ensemble
averaging needs to be performed.  This averaging leads to a different
time constant $T_2^*(\le T_2)$ to describe the width of the
magnetic resonance signal (the inhomogeneous broadening),
but it does not affect the longitudinal direction.  Note that $T_2$
(or $T_2^*$) describes the dephasing process ($T_2$ is often called
the dephasing time), and $T_1$ ($\ge T_2$) is the inelastic spin-flip
or spin-lattice relaxation time. Often $T_2$ is also called the
spin-spin relaxation time for reasons to be discussed below.  The
Bloch equations successfully describe dephasing and relaxation
phenomena in atoms, quantum optics, nuclear spins, and electron spins
in semiconductors \cite{slichter,Poole,meystre,Sham1}, even though 
microscopically it
may well be the case that only two time scales are not sufficient to
describe the dynamics, or that a continuous time description of the
single spin magnetization is only appropriate as an average behavior,
since the spin might be subjected to discrete random ``kicks''.  

Two particular solutions of the Bloch equations are of special
interest.  If there is no pure dephasing process (which we will
discuss below) in the system, $T_{1}=T_{2}$, i.e. all dephasing
arises from spin-flip inelastic scattering.  In this case the
magnetization rotates at the generalized Rabi flopping frequency
$R=\sqrt{R_{0}^{2}+\delta^{2}}$, and the equations for the
magnetization can be written in a succinct vectorial form:  
\begin{equation}
{\mathbf {\dot M}}=-\frac{1}{T_1}({\mathbf M}+{\mathbf {\hat z}})+
(\delta {\mathbf {\hat z}}-R_{0}{\mathbf {\hat x}})\times {\mathbf M}.
\label{equal_T}
\end{equation}

In general, $T_{1}\neq T_{2}$, and Eq.~(\ref{equal_T}) does not
apply.  Without the external tipping field, so that
$R_{0}=0$, the system undergoes a so-called free induction decay
\cite{slichter,Poole,meystre} if it is initially pushed out of
equilibrium: 
\begin{equation}
{\mathbf {\dot M}}=-\frac{1}{T_{2}}{\mathbf M_{\perp}}+\omega 
{\mathbf {\hat z}}\times {\mathbf M_{\perp}} -
\frac{1}{T_{1}}(M_{z}+1){\mathbf {\hat z}}. 
\end{equation}
The density matrix takes on the form of Boltzmann distribution at
large time ($t \gg T_{1}$) and finite temperature:
\begin{equation} \rho(t)= \left(
\begin{array}{cc}
(\rho_{\uparrow \uparrow}(0)-a_{0})e^{-\frac{t}{T_{1}}} +a_{0} 
& 
\rho_{\uparrow \downarrow}(0)e^{-i\omega t}e^{-\frac{t}{T_{2}}} 
\\
\rho_{\uparrow \downarrow}^{*}(0)e^{i\omega t}e^{-\frac{t}{T_{2}}} 
&  
(a_{0}-\rho_{\uparrow \uparrow}(0))e^{-\frac{t}{T_{1}}}+1-a_{0}
\end{array}
\right),
\label{eqrho2}
\end{equation}
where $a_{0}=e^{-\beta\mu_{e}B}/(e^{-\beta\mu_{e}B}+e^{\beta\mu_{e}B})$ 
is the thermal partition function at equilibrium with $\beta = (k_B
T)^{-1}$ as usual (where $T$ is the absolute temperature, {\it not} a
relaxation time).  

Regarding the decoherence of a single spin, we note that spin-flip
processes cause both population relaxation and dephasing,
contributing to both rates $1/T_{1}$ and $1/T_{2}$.   However, in a
real physical system the longitudinal and transverse directions are
often affected differently by the environment.  Indeed,  there exist
pure dephasing processes which affect only $T_{2}$ but not $T_{1}$. 
One example is the colliding molecules in an optically active
gaseous medium, where molecules constantly undergo collisions with
each other, some of them inelastic, but most of them elastic. During
an elastic collision, the single molecule energy spectrum changes due
to the presence of the other molecule nearby.  This  shift in energy
levels (particularly the two active levels) is dependent on the
details of the collision, and is thus a random variable, which we
refer to as $\delta \omega(t)$. Including this frequency shift the
differential equation (\ref{evo2}) for $\rho_{\uparrow \downarrow}$
becomes (here we are using the spin language to describe an orbital
TLS, with spin up and down  referring to the two levels)
\begin{equation} 
{\dot \rho}_{\uparrow \downarrow}(t)=-i(\omega +
\delta \omega(t))\rho_{\uparrow \downarrow}. 
\label{fluctrho}
\end{equation} 
Note that in the first order approximation $\rho_{\uparrow
\uparrow}$ is not affected since its equation of motion (\ref{evo1})
is independent of $\omega$.  $\delta\omega(t)$
is a random variable that averages to zero, $\langle \delta \omega
(t)\rangle=0$.  In the Markovian approximation (no memory effects in
the random variation of $\delta\omega(t)$) we have  
\begin{equation}
\langle\delta\omega(t)\delta\omega(t')\rangle=2\gamma_{ph}\;
\delta(t-t'). 
\label{2cor} 
\end{equation} 
This random fluctuation in energy level splitting of the TLS then
causes a pure dephasing effect:
\begin{equation}
\langle \rho_{\uparrow \downarrow}(t)\rangle=\langle \rho_{\uparrow 
\downarrow}(0)\rangle 
e^{-i\omega t}e^{-\gamma_{ph}t}.
\end{equation}
This pure dephasing only contributes to $T_2$, but not
to $T_1$.  Another well-known example of pure dephasing is the
dipolar spin-spin interaction in NMR, which produces effective local
magnetic field fluctuations and hence contributes essentially only to
$T_2$ (the corresponding
effect on $T_1$ is extremely small).  What is important for dephasing
is that some change in the state of the environment must occur due to
its interaction with the system---dephasing does not necessarily
require an explicit inelastic scattering process for the system,
although all inelastic scatterings necessarily produce dephasing.
In fact, as mentioned before, $T_2$ in the context of ESR and NMR is
often called the spin-spin relaxation time because the most important
intrinsic effect contributing to $1/T_2$ is the dipolar interaction
among various spins in the system, which, while transferring energy
among the spins themselves, does not lead to overall energy
relaxation from the total spin system.  By contrast, spin-lattice
interactions lead to energy relaxation (via spin-flip processes) from
the spin system to the lattice, and thus contribute to $T_1^{-1}$,
the spin-lattice relaxation rate.  We note in this context that $T_2$
sets the time scale for the spin system to achieve equilibrium within
itself whereas $T_1$ sets the time scale for the global thermodynamic
equilibrium between the spin system and the lattice.  It should be
emphasized that all inelastic processes contributing to $T_1$ also
automatically lead to dephasing, but in many circumstances there may
be additional dephasing processes (e.g. dipolar spin-spin coupling in
NMR and ESR) which contribute only to $T_2$ (and {\it not} to $T_1$),
and therefore $T_1 \ge T_2$ in general.

\section{Measuring $T_{1}$, $T_{2}$, and $T_{2}^{*}$ of electron spins 
in semiconductors}

In metals and doped semiconductors there are three major spin
relaxation mechanisms for conduction electrons \cite{optor,jaro}
except at the lowest temperatures, where relaxation times are
extremely long and dipolar and nuclear coupling become
comparatively important.  In the Elliot-Yafet mechanism,
spin-independent interactions of electrons with impurities,
boundaries, interfaces, and phonons connect up and down spins through
spin-orbit coupling. In the Dyakonov-Perel' mechanism, the splitting
of spin up and down conduction bands due to lack of inversion
symmetry (as in III-V semiconductors which have the zinc-blende
lattice structure) acts as an effective momentum dependent magnetic
field ${\mathbf B}({\mathbf k})$.  An electron with momentum ${\bf
k}$ and spin ${\bf S}$ precesses in this effective field ${\bf
B}({\bf k})$ and loses its spin memory.  As this electron is
scattered into a different ${\bf k}$ state, its spin will start to
precess around the new effective field.  This constant change of
effective magnetic field actually impedes the electron spin
relaxation, so that the spin relaxation time is inversely
proportional to the momentum relaxation time in this mechanism.  The
third mechanism, the Bir-Aronov-Pikus mechanism, is given by the
exchange interaction between electrons and holes.  Electronic spins
move in an effective field produced by the hole spins, and relaxation
takes place when hole spins change in a rate much faster than the
electron precession frequency.  A more detailed review of these
mechanisms and their experimental observations in metals and
semiconductors have been discussed in reference \cite{jaro}.  We only
mention here that in general electron spin relaxation in GaAs is weak
due to its relatively weak conduction band spin-orbit coupling---a
simple matrix element estimate indicates that the spin relaxation
time is of the order of 10-100 ns at low temperatures ($T \sim
4$ K), and one should approximately have $T_1 \sim T_2$ in high
quality GaAs.  Such a ``long" relaxation time ($T_2$) has recently
been directly measured in ESR experiments \cite{ESR}.  We should
emphasize that this relaxation time ($\sim 10-100$ ns) is ``long"
only in a relative sense compared with electron-electron scattering
times ($\sim$ fs) or momentum relaxation times ($\sim$ ps)---electron
spin relaxation in GaAs should be rather slow due to very weak
conduction band spin-orbit coupling, and the observed relaxation
times are not long in any absolute sense.  In fact, low temperature
electron spin relaxation times in metals are also very long ($\sim$
$\mu$s) \cite{jaro}.

Coherent manipulation of electron spins and the study of electron spin
relaxation have a long history going back to the first ESR experiments
half a century ago \cite{Poole}, and ESR remains a
key technique to determine electron spin relaxation and study
electron spin dynamics \cite{ESR}.  Below we focus on
some of the recent experiments that employ alternative
optical approaches to study electron spins in semiconductors,
especially on optical orientation and related measurements of spin
relaxation, which may have particular relevance to solid state
quantum computation, where externally controlled optical pulses may
be used to coherently manipulate electron spin dynamics in
zinc-blende semiconductor structures.

\subsection{Bulk III-V semiconductors}

The method of optical orientation can be used to measure 
spin relaxation ($T_{1}$) in bulk III-V semiconductors like
GaAs \cite{optor}.  
%
%It generates spin polarized electrons
%and holes in the conduction and valence bands respectively by shining
%circularly polarized light.  Photons of right or left polarization
%carry $+\hbar$ and $-\hbar$ of angular momentum in the direction of
%the light beam.  Since this quantity is conserved the total spin
%of the electron and hole will be equal to the photon's angular
%momentum. Therefore by 
%
Shining circularly polarized light on a semiconductor generates
polarized conduction electrons with spin oriented in the direction
of the light beam.  These spins will relax until
the electrons recombine with holes in a typical time $\tau_{p}$ when a
photon is emitted.  If $\tau_{p}\ll T_{1}$, then the recombination
photon will also be circularly polarized.  Since $\tau_{p}$ can be
extracted by other techniques one can
measure $T_{1}$ by studying the polarization of the recombination
luminescence.  If a magnetic field is applied
perpendicular to the spin orientation of the electrons, the
recombination radiation is less polarized (Hanl\'{e} effect).  This
effect can be used to measure both $\tau_{p}$ and
$T_{1}$\cite{optor}.  Typical values for $T_{1}$ in p-doped GaAs are
$10$ ps to $1$ ns, depending on the sample temperature and impurity
concentration.  Thus, the hole spin relaxation rate is relatively
fast in GaAs, consistent with its strong valence band spin-orbit
coupling.

Optical measurements have recently been used to
measure $T_{2}^{*}$ in n-doped samples of bulk GaAs, using
pump-probe and Faraday rotation techniques \cite{kikkawa}.  
A circularly polarized laser pulse is used to pump electrons to the
conduction band.  The holes left behind by the excited electrons
quickly lose their spin polarization (due to strong valence band
spin-orbit coupling for holes) and are annihilated by the
doped electrons in the first 100 ps after the pump pulse,
substantially reducing the hole-related spin relaxation
(Bir-Aronov-Pikus mechanism).  A linearly polarized laser pulse is
then used to probe the remaining electron spins a time $\Delta t$
after the pump pulse.  When this pulse scatters off the partially
polarized electrons, its linear polarization is rotated by an angle
$\theta$ proportional to the projection of the spin magnetization in
the direction of the laser beam, a phenomenon known as Faraday
rotation. By using pump and probe pulses of about 100 fs in duration
respectively, the time resolved Faraday rotation technique can
reveal the time dependence of the magnetization of the spins in the x
direction,   
\begin{equation} M_{x}=M_{x}(\Delta t=0) \;
e^{-\frac{\Delta t}{T_{2}^{*}}}\cos (g\mu_{e}B\Delta t /\hbar) \,,
\end{equation} 
with $\Delta t$ being the time delay between pump and
probe pulses. At $B=1$ T, $T=5$ K, and doping concentration
$n=10^{16}cm^{-3}$, a spin decoherence time $T_{2}^{*}$ of
approximately 10 ns was measured \cite{kikkawa}. 
Furthermore, $T_{2}^{*}=130$ ns is obtained at $B=0$, which is one
order of magnitude larger than the value at $B=1$ T.
Since the spins are not precessing at $B=0$, one could argue
that in this particular case $T_{2}^{*}=T_{2}=T_{1}=130$ ns, which is
the decay of the magnetization in all direction since there is no
magnetic field to break the symmetry, hence the increase by one order
of magnitude (note that without an external magnetic field,
the g factor inhomogeneities discussed in \cite{kikkawa,gupta} will
not contribute to $T_{2}^{*}$).  This value of the relaxation time is
consistent with the direct ESR measurement of
the linewidth to be around 50 MHz in GaAs heterostructure \cite{ESR}.

Time-resolved 
Faraday rotation was also used to study spin precession in an ensemble
of chemically synthesized CdSe quantum dots \cite{gupta}. 
$T_{2}^{Inh}$ (which contains averaging not only for the many
electrons in one quantum dot (QD), but also over many quantum dots)
was measured
ranging from $3$ ns at $B=0$ to less than $100$ ps at $B=4$ T.  It was
suggested that this strong field dependence comes from the
fact that the quantum dots have varying g factors ranging between
1.1 and 1.7.  To measure the decoherence time $T_{2}^{*}$ of the many
electrons in a single quantum dot one could in principle use the
spin echo technique to constantly realign the spins and observe the
time decay of the transverse magnetization, hence extracting
$T_{2}^{*}$ from the data of an ensemble of quantum dots.  This
experiment has not yet been performed, although the tipping technique
required by spin echo type of experiments has recently been
demonstrated \cite{gupta1}.

\subsection{Single spin in a quantum dot}

Based on the assumption that an electron spin
possesses long coherence times ($T_{1}$, $T_{2}$ much greater
than the time to perform qubit operations, see below), a
spin-based quantum dot quantum computer (QDQC) has been proposed
\cite{LD}. It is thus desirable to have estimates of $T_{1}$ and
$T_{2}$ for single electron spins in a quantum dot, and to confirm the
estimates experimentally.  It is important to mention in this context
that the single spin decoherence times should be longer than the
macroscopic (averaged over many spins) values of $T_1$ and $T_2$,
which should thus serve as an upper limit.

Phonon-assisted spin flip rates due to spin-orbit coupling in a
single electron GaAs QD has been calculated \cite{khaetskii}.  It was
pointed out that, due to wave function localization, the spin orbit
relaxation mechanisms discussed above for a free electron
(Elliot-Yafet, Dyakonov-Perel') are strongly suppressed in a QD,
giving a long spin flip time: $T_{1}\approx 1$ ms for $B=1$ T and
$T=0$ K.  It was further noticed that spin relaxation is dominated by
the Elliot-Yafet mechanism, which yields $T_{1}\propto B^{-5}$ for
transitions between Zeeman sublevels in a one electron QD. 

These calculations are consistent with recent transport measurements 
\cite{fujisawa}.  Pulses
of current were injected into a QD coupled to leads in the Coulomb
Blockade regime, where the decay rate from excited states can be
measured by analyzing the decay time of transient currents generated
by the pulses.  The results indicate that, for $T=150$ mK and $B=0-2$
T, spin relaxation times ($T_{1}$) in a many-electron QD (less than
50 electrons) are longer than at least a few $\mu$s.  This is
encouraging from the perspective of the spin-based solid state QC
architecture where spin relaxation times of $\mu$s or longer are
most likely necessary for large scale QC operation.

ESR combined with transport techniques in principle could be used
to probe $T_{2}$ in a QD in the Coulomb Blockade regime.  It
has recently been proposed \cite{engel} that by applying an AC pump
field to a single electron QD subjected to a magnetic field, the
stationary current through this QD will exhibit a peak as a function
of the pump frequency, whose width will yield a lower bound on
$T_{2}$.  This is in essence an ESR measurement with a transport
readout.

It is desirable that $T_2$ for an electron in a single QD is a factor
of $10^{4}$ or so greater than the typical gating time in a QDQC
\cite{nielsen}.  For $B=1$ T, the Zeeman splitting in a QD is about
$0.03$ meV,  which yields $100$ ps for the precession time of one
spin, which can be used as the one qubit gate (the two qubit gate
time is lower, $\hbar/J\sim 50$ ps for $J\sim 0.1$ meV).  Therefore
for quantum error correction to be performed reliably, $T_{2}$  for
the trapped electron spin needs to be on the $\mu$s time scale, which
may very well be the case at low enough temperatures in a single QD. 
We note that the existing experimental estimates of free electron
spin relaxation time $T_2$ (or $T_2^*$) in GaAs (for $T = 1$--4 K) is
around 10--100 ns, which is obviously a lower bound since one expects
on rather general ground the spin relaxation time of individual
electrons confined in QD structures to be longer.  Much of the
current optimism in the feasibility of spin-based QDQC architecture
arises from the expected long electron spin relaxation times (many
$\mu$s or longer at low temperatures) of GaAs electrons confined in
high quality QDs.

\section{Decoherence in a quantum dot quantum computer}

Defining a TLS involves choosing two levels that are well isolated
from the rest of the degrees of freedom in the overall system.
In a solid state system, where the density of states is generally high
and where long range electromagnetic interaction typically couples
all the charged entities, isolating two levels is particularly
difficult.  In a spin-based quantum computer, the spin
up and down states are chosen as the two states of a qubit.  It is
then crucial to explore all relevant degrees of freedom in the solid
state environment that couple to the electron spins and determine
whether they are sufficiently weak (or can be made so by suitable
means) for a quantum computer to work.

Trapped in an ideal quantum dot in a perfect heterostructure, an
electron is sitting at the bottom of the conduction band, and there
should be no other electrons or holes nearby in order for the
trapped electron to be considered as a candidate for qubit.  Thus the
DP and BAP spin relaxation mechanisms should not be important.  The
major environmental influences come from spin-orbit coupling
(therefore phonons) and nuclear spins in the material, assuming
dipolar coupling to other electron spins to be weak; otherwise
dipolar coupling to other spins has to be accounted for as well.  The
experimental temperature can be lowered so that no resonant phonon
(producing decoherence) is present in the system, and nuclear spins
need to be controlled via various means.  However, these controls are
not sufficient because imperfections (e.g. impurities, defects, etc.)
are facts of life in artificially created QD structures.  Thus there
are additional factors that need to be considered such as imperfect
interfaces and nearby metallic gates as well as the associated charge
diffusion and charge traps.  Boundary scattering can probably be
sufficiently suppressed by enhancing interface material quality and
by making the quantum dots small.  On the other hand, it is much
harder to control charge transfers between the metallic gates and the
heterostructure in which the quantum dots are located.  Since the
gates cannot be too far away from the dots (in order to produce small
dots), any charge diffusion or trapping can lead to important (and
often unknown) environmental effects.  For example, if a stray
electron is trapped too close to a quantum dot, there may be an
exchange interaction between the electrons, which would cause
complete loss of spin information in the quantum dot, destroying
qubit operation.

When we bring two quantum dots close to each other in order to perform
exchange gate operations required for quantum computing \cite{LD},
there are additional external influences on the system.  One concern
is whether Heisenberg exchange Hamiltonian is a complete description
of the low energy dynamics of a double dot.  The higher excited states
should certainly be eliminated (i.e. pushed to high energy) from the
system, which we will discuss further below.  Inherent spin-orbit
coupling in 2D quantum dots leads to an anisotropic exchange, whose
effects have to be controlled and, if necessary, corrected
\cite{Kavokin,BSD}.  It is self-evident that the environmental
influences for both single and double quantum dot structures will
have to be carefully studied before a functional QDQC (even
with just a few qubits) can be fabricated.  

For a QC to work, we need to understand not only its full environment,
but also how to best control it.  To operate a QC, we inevitably
need to manipulate the spins (or other form of qubits) using external
means: gate voltages, applied magnetic fields, light, microwave, etc.
When we introduce these external influences, we inevitably introduce
unwanted perturbations.  These can be called gate errors.  Since these
external perturbations necessarily
involve some leakage of the system into its environment,
they can also be regarded as a form of decoherence.

For example, in performing exchange gates in spin-based QDQC, the
potential barrier between two neighboring quantum dots needs to be
lowered to allow the electron wavefunctions to overlap and thus
enabling the exchange interaction to take effect.  However, the
mixing of the two electrons naturally leads to the possibility of
exciting the previously frozen orbital degrees of freedom, especially
if the mixing happens too fast.  It has been pointed out that
adiabatic operation of the exchange gate can help suppress the errors
caused by the state mixing \cite{BLD,SLM}.  Essentially, if the
two-qubit system is initially prepared as a superposition of the
ground singlet and triplet states, as long as the adiabatic condition
is satisfied, the system will remain in the sub-space spanned by the
ground singlet and triplet states, no matter how these states change
during the barrier variation.  Of course, the adiabatic condition can
never be satisfied exactly in any real situation, therefore the
important issue is to estimate the amount of mixing of the higher
energy states under realistic conditions.
We have recently done a calculation \cite{HDad} of the 
time evolution of a two-electron double dot system when the central
barrier between the dots is varied, so that we can quantitatively
determine the adiabatic condition that a spin-based QDQC has to satisfy.  

Under the adiabatic approximation, a quantum
state can be expanded in the instantaneous basis of the system:
$\psi(t) = \sum_i c_i(t) u_i(t)$, and the instantaneous basis states
satisfy $H(t) u_i(t) = E_i(t) u_i(t)$.  Thus the Schr\"{o}dinger
equation takes the form:
\begin{equation}
\frac{\partial c_k}{\partial t} = \sum_{i \neq k}^{N} \frac{c_i}{E_k
- E_i} \langle k |\frac{\partial H}{\partial t}| i\rangle \, 
\exp \left\{\frac{1}{i\hbar} \int_{-\infty}^t (E_i - E_k) d\tau 
\right\} \,.
\end{equation}
Here $N$ is the total number of basis states involved.  To perform
a simulation of the dynamics of this system, we need to know
all the eigenstates and eigenenergies at all time.  Since
the eigenstates and energies vary relatively slowly with the change
in the Hamiltonian, specifically the change in the barrier height
between the two dots in our case, we did the eigensystem calculation
on a finite grid, then obtained the rest of the information through
interpolation.  Our results show \cite{HDad} that for a typical
configuration \cite{HD} of the double dot system for quantum
computing, the leakage rate would be reasonably small ($<
10^{-6}$) for gate operation times longer than 50 ps.  Thus adiabatic
condition is not overly stringent and should not hamper the operation
of a QDQC.  Note that the optimum value of the gate
operation time is constrained from below by the adiabatic condition
(i.e. gates should not be too ``fast") and from above by the spin
relaxation time (i.e. gates should be much faster than the typical
spin relaxation time).  Our most optimistic reasonable estimate
of the applicable range of gating time for spin-based QDQC
operations is 50 ps to 1 ns, which is not an unrealistic operation
regime. 

As the exchange coupling $J$ (the singlet-triplet splitting) is tuned
\cite{HD} by changing external gate voltage in a QDQC,
thermal fluctuations (or any other types of fluctuations) in the gate
voltage will lead to fluctuations in $J$, thus causing phase errors
in the exchange-based swap gate which is crucial for two-qubit
operations.  We have estimated this error by assuming a simple
thermal (white) noise \cite{HD}.  More specifically, we assume
$J=f(V)$ where  $V$ is the gate voltage that controls the value of
$J$.  Around any particular value $V_0$, $J$ can be expressed as
$J(V) = J(V_0) + \left. f^{\prime}(V) \right|_{V_0} (V-V_0)$.  During
a swap gate between two quantum dots, the phase of the electronic 
spin wavefunction evolves as $\phi = \int_0^t J(\tau) d\tau /\hbar$.
Thus the fluctuation in the phase $\phi$ is 
\begin{eqnarray}
\langle \delta \phi^2 \rangle & = & \langle \phi^2 \rangle
- \langle \phi \rangle^2 \ = \ \frac{1}{\hbar^2} \int_0^t \int_0^t
\langle \delta  J(\tau_1) \ \delta J(\tau_2) \rangle d\tau_1 d\tau_2 
\nonumber \\
& \sim & \int_0^t \int_0^t 
\frac{[f^{\prime}(\bar{V})]^2}{\hbar^2}
\langle \delta V(\tau_1) \ \delta V(\tau_2) \rangle 
d\tau_1 d\tau_2 \,.
\end{eqnarray}
Using Nyquist theorem $\langle \delta V(\tau_1) \ \delta V(\tau_2)
\rangle = 4 R k_B T \delta(\tau_1-\tau_2)$,
we obtain the approximate expression for the phase fluctuation:
\begin{equation}
\langle \delta \phi^2 \rangle
\sim 4 R k_B T \alpha^2 t /\hbar^2 \,,
\end{equation}
where $\alpha$ is the upper bound of $|f^{\prime}(\bar{V})|$.
Assuming the swap gate is performed at 1 K
(since $J$ is in the order of 0.1 meV $\sim$ 1 K, the
experimental temperature cannot really be much higher than 1 K),
and the transmission line connecting the gate to the
surrounding cryogenic bath has an impedance
of 50 ohms, the rate for phase fluctuation $\langle \delta \phi^2
\rangle/t$ is about 3.2 MHz. The phase error accrued during a swap
gate is then about  0.06\%.  This is quite a small error which is
of the same order of magnitude as the theoretical tolerance of the 
currently available quantum error correction codes.
To further lower this error rate, one can lower the
experimental temperature and set up the gate architecture in such
a way as to decrease the sensitivity of the exchange coupling $J$
on the gate voltage $V$ in the operation of a QDQC.

Another possible error in the two-qubit operations of the QDQC
architecture is caused by inhomogeneous magnetic
fields \cite{HDD,DHD}.  Such a field may come from magnetic impurities
or unwanted currents away from the structure.  Magnetic field affects
both orbital and spin parts of the electron wavefunction.  The orbital
effect is accounted for by adjusted exchange coupling $J$, while the
spin effect is accounted for through Zeeman coupling terms:
\begin{eqnarray}
H_s & = & J({\bf B}) {\bf S}_1 \cdot {\bf S}_2 + \gamma_1 S_{1z} 
+ \gamma_2 S_{2z} \,,
\label{eq:spin-Hamiltonian} 
\end{eqnarray}
where ${\bf S}_1$ and ${\bf S}_2$ refer to the spins of the two
electrons, $J({\bf B})$ is the exchange coupling (singlet-triplet
splitting), and $\gamma_1$ and $\gamma_2$ are the effective strength
of the Zeeman coupling in the two quantum dots.  In an inhomogeneous
field, $\gamma_1 \neq \gamma_2$, so that the Zeeman terms do not
commute with the exchange term in the Hamiltonian 
(\ref{eq:spin-Hamiltonian}).  We have done a detailed analysis
\cite{HDD,DHD} on
how to achieve swap with such a Hamiltonian, and found that there is
at the minimum an error proportional to the square of field
inhomogeneity in the swap.  For example, if the initial state of the
two electron spin is $|\phi(0)\rangle=|\!\uparrow \downarrow \rangle$,
the density matrix of the first spin after the optimal swap
is
\begin{equation}
\left. \rho_1 \right|_{e^{i\theta}=-1}
=\frac{1}{1+x^2}|\!\downarrow\rangle \langle \downarrow\!|
+ \frac{x^2}{1+x^2}|\!\uparrow\rangle \langle \uparrow\!| \,,
\label{eq:swap_error}
\end{equation}
where $x=\delta/2J=(\gamma_1-\gamma_2)/2J$.  In other words, 
the first spin can never exactly acquire the state ($|\!\downarrow 
\rangle$) of the second spin.  Its state will remain mixed and
the smallest error from an exact swap is $x^2/(1+x^2)$,
which needs to be corrected.  We have estimated \cite{HDD} that in
GaAs a Bohr magneton can lead to an error in the order of $10^{-6}$,
which is within the capability of currently available quantum
error correction schemes.

In concluding this section, we mention that many of the techniques
developed over the last fifty years in the context of NMR and ESR
studies should be useful in controlling decoherence (and carrying out
error corrections) in spin-based QDQC operations.  These include the
spin echo and refocusing techniques.  In fact, our best estimates for
the electron spin relaxation time in GaAs quantum dots
($\sim \mu s$), the exchange coupling ($\sim$ 0.1 meV), and the
gating time (ps to ns) makes us guardedly optimistic that a
spin-based QDQC architecture may very well be developed in
the future.  It should, however, be kept in mind in this context that
QC architectures \cite{nielsen} based on atomic physics (e.g. 
trapped ions) and liquid state NMR have extremely long natural
decoherence times ($T_2 \sim \mu$s-ms; $T_1 \sim$ minutes-hours)
because of the extremely weak environmental coupling in these
systems.  The main problem in these architectures is not decoherence,
but scaling up to more than a few (2-10) qubits which should be
relatively easy in semiconductor solid state systems \cite{Kane}.

\section{Mesoscopic decoherence}

Finally, we provide a very brief discussion on the issue of
electronic decoherence in mesoscopic electronic materials (e.g.
metals, semiconductors).  Our discussion is necessarily brief (done
only in the context of quantum computing), and certainly will not do
justice to the vast literature and the great deal of current activity
in the subject.  We provide this discussion only for the sake of
completeness, restricting ourselves entirely to rather elementary
considerations.

Electronic decoherence in mesoscopic materials is often characterized
by dephasing (or the phase relaxation) time $\tau_{\phi}$ which
shows up in electronic phase coherent processes (i.e. quantum
interference phenomena) such as weak localization, conductance
fluctuations, and various quantum interference oscillations (i.e.
$h/e$ and $h/2e$ oscillations in connected structures).  In general,
each quantum coherent phenomenon may involve a slightly different
(but closely related) definition of $\tau_{\phi}$, but in all cases 
$\tau_{\phi}$ indicates the typical dephasing time over which quantum
interference memory is lost in the system \cite{AAK}.  The dephasing
time $\tau_{\phi}$ in the many-body electronic system is similar in
spirit to the dephasing time $T_2$ in the TLS case
although, in contrast to TLS dynamics, the concept of a
transverse relaxation time (as $T_2$ is in the TLS case) is not
germane to the mesoscopic problem.  The inelastic scattering (or
relaxation) time $\tau_i$ (sometimes also called quasiparticle
lifetime) in the electronic many-body system corresponds loosely to
the inelastic spin-flip relaxation time $T_1$ in the TLS problem. 
Actually, the loose analogy between $T_2$ and $\tau_{\phi}$ on the
one hand and $T_1$ and $\tau_{i}$ on the other hand could be further
elucidated.  Just as one of the main contributions to $1/T_2$ at low
temperatures is the spin-spin (dipolar) interaction, the main
contribution to $1/\tau_{\phi}$ at low temperatures is the
electron-electron interaction.  Similarly, if one restricts to energy
loss from the whole electron system (and not just the individual
electrons), then electron-phonon interaction is the main inelastic
scattering mechanism in electronic systems, just as spin-lattice
relaxation is the inelastic spin-flipping mechanism in $T_1$.

While $\tau_i$ is an inelastic lifetime related to the inverse of the
one-electron self-energy, $\tau_{\phi}$ relates directly to the
dephasing process.  They are closely connected, but {\it not}
necessarily the same, particularly at low temperatures and in low
dimensional systems.  In general, elastic scattering by impurities
(where the impurities are considered static and immobile) does not
{\it directly} contribute to either $\tau_i$ or $\tau_{\phi}$.  The
presence of impurity scattering does, however, have strong {\it
indirect} influence on both $\tau_i$ and $\tau_{\phi}$, since the
electronic motion in mesoscopic systems is diffusive (with a finite
transport mean free path) in the presence of impurity scattering. 
For example, in a 3D metallic system the electron-electron scattering
contribution (which is expected to dominate at low temperatures
where the phonons are frozen out) to $\tau_i$ is $\tau_i^{-1} \sim
T^2$ for a ballistic system, and $\tau_{\phi}^{-1} \,, \tau_i^{-1} 
\sim T^{3/2}$
in a diffusive system.  Note that in a diffusive 3D system
$\tau_{\phi} \sim \tau_i$ for electron-electron scattering (this is
also true for electron-phonon scattering which is important at higher
temperatures).  This approximate equality of $\tau_i$ and
$\tau_{\phi}$ for 3D systems follows primarily from the fact that
large energy transfer ($\sim k_B T$) scatterings dominate both 
$\tau_i$ and $\tau_{\phi}$ in 3D systems.  In lower dimensional
systems, however, this is not true in general, and we can have 
$\tau_i \gg \tau_{\phi}$ with $\tau_{\phi}^{-1}$ being dominated by
very small energy scattering processes which in general do not much
affect $\tau_i^{-1}$.  The calculated temperature dependence of 
$\tau_i$ and $\tau_{\phi}$ due to electron-electron scattering in
2D diffusive systems are, however, the same up to a logarithmic
correction: $\tau_i^{-1} \sim T \ln T\,, \tau_{\phi}^{-1} \sim T$.  In
1D, on the other hand, $\tau_{\phi}^{-1} \sim T^{2/3}$ and
$\tau_i^{-1} \sim T^{1/2}$.  For electron-phonon scattering one
typically finds \cite{BD} $\tau_{e-ph} \approx \tau_{\phi} \sim
T^{-p}$ with $p \approx $1--4 depending on systems and
dimensionalities.  Usually electron-electron scattering dominates
$\tau_{\phi}$ for $T < 10$ K and electron-phonon scattering
dominates at higher temperatures.  It may be worthwhile to point
out that just as the existence of $T_2$ (with $1/T_2 \neq 0$) is
essential in defining a spin temperature $T_s$ (through spin-spin
interaction) in ESR and NMR measurements \cite{slichter,Poole},
electron-electron scattering is essential in defining an electron
temperature $T_e$ in an excited electron gas.  The final equilibrium
to the lattice temperature $T_L$ (i.e. $T_s \rightarrow T_L$ or
$T_e \rightarrow T_L$) is achieved in both cases through the
interaction with the lattice (i.e. $1/T_1 \neq 0$ or $1/\tau_{e-ph}
\neq 0$).

The strong temperature dependence of 
$\tau_{\phi}$ ($\sim T^{-3/2}$ in 3D, $T^{-1}$ in 2D, and $T^{-2/3}$
in 1D) at low temperatures is only observed over a rather limited
range of temperature in experiments
where the measured $\tau_{\phi}$ in many different (particularly, low
dimensional mesoscopic) systems always seems to saturate at low
enough temperature and cross over to a constant value of the order
of 1 ns (with the crossover temperature of the order of 100 mK)
\cite{Webb}.  Neither the saturation temperature nor the saturated
value of $\tau_{\phi}$ seems to have any obvious universal behavior. 
Whether this low temperature $\tau_{\phi}$ saturation has any
fundamental significance or not is currently being debated in the
literature.  We have nothing to add to this controversy other than to
point out that even in the much simpler problem of NMR, actual
quantitative calculations of $T_2$ are rather difficult because so
many different processes could cause dephasing at low  temperatures. 
In the $\tau_{\phi}$ problem, such dephasing processes include, for
example, electron heating, magnetic impurities, background radiation,
unknown inelastic channels at very low energies, finite system
sizes, and actual slight movements of the impurity atoms which are
considered to be static in the theory.  Indeed, at low enough
temperatures, where the dephasing time is already relatively long,
any weak coupling to the environment will cause (weak) decoherence
and lead to apparent saturation of $\tau_{\phi}$ at finite effective
electron temperatures.

\section{Conclusion}

We conclude by emphasizing that understanding dephasing and
decoherence in specific solid state quantum computer architectures is
crucial in the development of successful qubit operations in scalable
solid state systems.  For example, the Rabi oscillations observed
\cite{Nakamura} in the superconducting Cooper pair box system seem to
have a very short dephasing timeof the order of nanoseconds only. 
Currently there is no theoretical understanding of such a short
dephasing time in this system, although various electrical noise
sources in the circuit could be one possible problem.  Similarly, the
anticipated flux qubitstates in the Josephson junction SQUID system
\cite{Schon,SQUID} have not yet demonstrated Rabi oscillations,
perhaps because of strong decoherence or for other fundamental
reasons.  In semiconductor quantum dot based spin qubit systems the
decoherence time ($T_2$ or $T_2^*$ depending on the specific
experiment) is experimentally measured to be in the 10 ns regime
\cite{gupta}, which is far too short for any effective quantum
computation operations---$T_2$ in the range of $\mu$s to ms will be
required for useful quantum computation.  Theoretical estimates
\cite{jaro,FD,JAP,Flatte} for relaxation times tend to concentrate on
$T_1$, which is a conceptually simpler quantity to compute (being an
inelastic relaxation time)---$T_2$ calculations must take into
account the details of the specific experimental setup.  In the
context of TLS dynamics, it is important to emphasize that while
$T_1$ refers to the lifetime of the excited $|1\rangle$ state (i.e.
$T_1$ is the time over which the excited state $|1\rangle$ decays to
the ground state $|0\rangle$ through the inelastic emission of
photons, phonons, magnons, etc.), $T_2$ refers to the coherence of
the superposition state $(|0\rangle+|1\rangle)/\sqrt{2}$ between the
ground and the excited state and as such depends on the precise phase
relationship between $|0\rangle$ snd $|1\rangle$ in a specific
experimental setup where all the details of the experimental
circuitry may matter in a nontrivial manner.  This makes $T_2$ ($\le
T_1$) intrinsically more intractable than $T_1$ from a theoretical
standpoint.  We point out that quantum computation is possible in
principle because single qubit quantum dephasing, while often being
continuous in actual experimental situations, can always be
equivalently represented as a discrete phase flipping process. 
Understanding dephasing in solid state quantum computer architectures
(or in general, in solid state systems) remains one of the important
problems in condensed matter physics.

\section*{Acknowledgement}

This work is supported by ARDA, DARPA, and ONR.

\end{document}